\documentclass[usenatbib,usegraphicx,usedcolumn]{mn2e}
\usepackage{times,amssymb,amsmath,mathrsfs}
\usepackage{relsize}
\usepackage[usenames,dvipsnames]{color}

\usepackage[backref,breaklinks,colorlinks,citecolor=blue]{hyperref}
\usepackage{rotating}
\usepackage{multirow}

\title[Integer Lattice]{Integer Lattice Dynamics for Vlasov-Poisson}
\author[P. Mocz et. al.]{Philip Mocz$^{1}$\thanks{E-mail: pmocz@cfa.harvard.edu (PM)} and 
Sauro Succi$^{2,3}$ \\
$^{1}$Harvard-Smithsonian Center for Astrophysics, 60 Garden Street, Cambridge, MA 02138, USA \\
$^{2}$Istituto per le Applicazioni del Calcolo, CNR, Viale del Policlinico 137, I-00161, Roma, Italy \\
$^{3}$Institute of Applied Computational Science, Harvard School of Engineering and Applied Sciences, Northwest B162, 52 Oxford Street, Cambridge, MA 02138, USA\\
}
\begin{document}

\date{MNRAS, xxx 2016}

\pagerange{\pageref{firstpage}--\pageref{lastpage}} \pubyear{2016}

\maketitle

\label{firstpage}

\begin{abstract} 
We revisit the integer lattice (IL) method to numerically solve the Vlasov-Poisson equations, and show that a slight variant of the method is a very easy, viable, and efficient numerical approach to study the dynamics of self-gravitating, collisionless systems. The distribution function lives in a discretized lattice phase-space, and each time-step in the simulation corresponds to a simple permutation of the lattice sites. Hence, the method is Lagrangian, conservative, and fully time-reversible. IL complements other existing methods, such as $N$-body/particle mesh (computationally efficient, but affected by Monte-Carlo sampling noise and two-body relaxation) and finite volume (FV) direct integration schemes (expensive, accurate but diffusive). We also present improvements to the FV scheme, using a moving mesh approach inspired by IL, to reduce numerical diffusion and the time-step criterion. Being a direct integration scheme like FV, IL is memory limited (memory requirement for a full 3D problem scales as $N^6$, where $N$ is the resolution per linear phase-space dimension). However, we describe a new technique for achieving $N^4$ scaling. The method offers promise for investigating the full 6D phase-space of collisionless systems of stars and dark matter.
\end{abstract}

\begin{keywords}
gravitation -- methods: numerical -- stars: kinematics and dynamics -- galaxies: kinematics and dynamics -- cosmology: dark matter

\end{keywords}

\begin{figure*}
\centering
\includegraphics[width=0.97\textwidth]{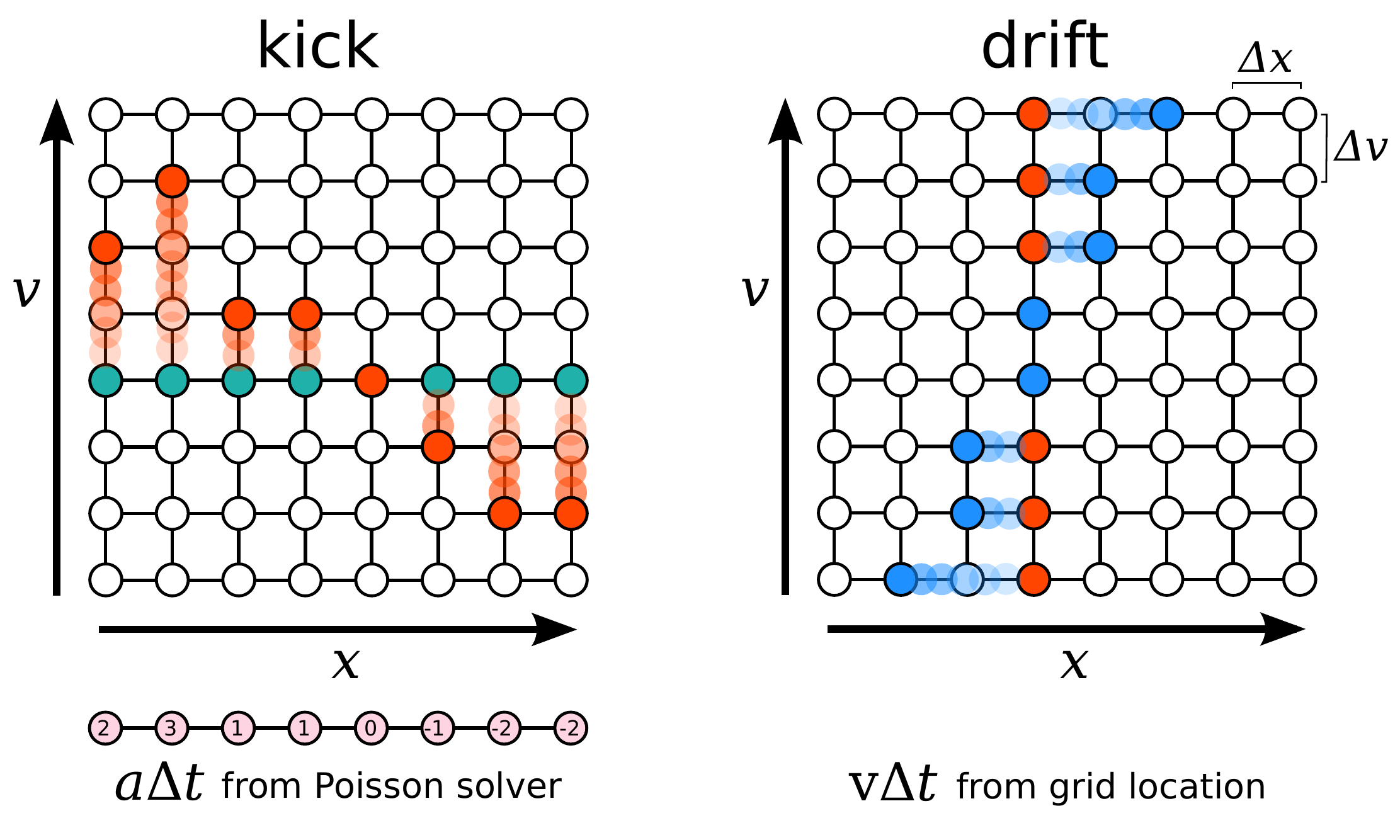}
\caption{Depiction of the IL algorithm. Each time step $\Delta t$ consists of a `kick' step due to gravitational acceleration, followed by a `drift' step due to advection. The accelerations $a$ are calculated from a Poisson solver and the $a\Delta t$ are rounded to the nearest integer to ensure the particles land on a lattice site.}
\label{fig:algorithm}
\end{figure*}

\section{Introduction}\label{sec:intro}

Two decades ago, \cite{1995MNRAS.276..467S} introduced an integer lattice (IL) method for the numerical simulation of collisionless stellar dynamics. 
The method solves the Vlasov-Poisson equation by shifting particles around on a lattice representing phase-space.
The method has several advantages: it is symplectic, Lagrangian, conservative, non-diffusive, and fully time-reversible. Additionally, the method is very easy to implement. In each time step, forces are rounded to the nearest integer so that particles always advect to a lattice site. 
The method has the unique advantage among other methods of realizing the Poincar\'e recurrence theorem: that is, with the IL method the system returns to its initial state after a sufficiently long, but finite time.
The IL method of \cite{1995MNRAS.276..467S} has two main drawbacks. First, as a direct integration scheme, the method is memory limited: a 3D problem requires storing information in a 6D phase-space lattice, meaning that the memory scales as $N^6$, where $N$ is the resolution per linear phase-space dimension. Second, the method is affected by, what we call, granular ``lattice noise'', due to rounding accelerations to the nearest lattice spacing. Thus the method has no formally derived order of convergence. However, it will still recover the solution in the limit of high resolution. 
\cite{1995MNRAS.276..467S} demonstrated IL on some simple problems to obtain equilibria solutions and study their linear stability. However, since its presentation, the method has received relatively little attention in the numerical simulation community. Here, we revisit IL and compare it to other numerical methods, and show that it can be accurately used to simulate \textit{dynamics}. We make a simple but substantial modification to the time-step of the IL method, necessary to to achieve resolve the dynamical timescale accurately. Additionally, we describe a way to reduce the memory requirement from $N^6$ to $N^4$ (adding extra computational cost), which makes the method very competitive for future studies of full 6D phase-space problems.

The most practical current method for solving the collisionless Vlasov-Poisson equation is the $N$-body technique (e.g., \citealt{2001NewA....6...79S,2003ApJS..145....1B}).
Here, the distribution function is sampled by Lagrangian particles of equal mass. The method may be thought of as a Monte-Carlo sampling of phase-space. Large areas of phase-space remain unsampled due to the finite mass resolution of the method, which makes the method very efficient. The forces experienced by the particles, naively an $\mathcal{O}(N^2)$ calculation due to pair-wise interactions, may be computed efficiently using various techniques, such as particle-mesh (PM; particles are binned into cells to solve the Poisson equation) or tree-based/fast multipole expansion algorithms (particles are sorted into a hierarchy of groups, and the gravitational field can be computed by summing over multipole expansions of these groups). $N$-body is memory efficient, compared to fully sampling the phase-space. Adaptive/hierarchical time-stepping schemes can further reduce the computational cost of the method and allow for the simulation of a system with large dynamic range \citep{2001NewA....6...79S}. The method does come with disadvantages, however, including Monte Carlo noise and artificial two-body relaxations \citep{2013ApJ...762..116Y}.
Monte Carlo noise sometimes leads to a failure to resolve accurately instabilities or Landau damping, and two-body relaxation violates the collisionless property of the system (often this is controlled to some extent by introducing a smoothing length for gravitational interactions to minimize the effects of close encounters). \cite{2008MNRAS.385..236V,2011MNRAS.413.1419V} discuss ways to improve resolving fine-grained phase-space structure of $N$-body simulations by evaluating the geodesic deviation equation along the trajectories individual $N$-body
particles.

Certain systems with complicated velocity structures in phase-space require more accurate methods. One basic approach is the direct integration of the collisionless Boltzmann equation using finite-volume (FV) methods  \citep{2013ApJ...762..116Y}. The distribution function is a conserved quantity governed by a hyperbolic partial differential fluid equation, making FV very applicable to obtain accurate solutions that formally converge to the exact solution. The method is memory expensive, however, requiring memory that scales as $N_x^DN_v^D$, where $N_x$ is the number of resolution elements per spatial dimension, $N_v$ is the number of resolution elements per velocity dimension, and $D$ is the physical dimension of the problem. For a 3D problem (6D phase-space), with $N_x\sim N_v\sim N$, the memory scaling is $N^6$. Another limitation of the method is that it is not Lagrangian, and therefore it requires a much more strict time-step criterion for numerical stability due to the Courant-Friedrichs-Lewy (CFL) condition. Finally, the method is diffusive, which can erase phase-space structure.

Other, more sophisticated techniques are been developed to study the full phase-space density distribution function, which will not be discussed in detail in the present scope of the work. One such method is the `cloudy' Vlasov solver \citep{2005MNRAS.359..123A}, which uses local basis functions around Lagrangian points with elliptical support, and an occasional remap scheme to make the basis functions round again. Having complete sampling of phase-space, the method is again expensive. 
Another promising method is the `waterbag' method \citep{2008CNSNS..13...46C,2014MNRAS.441.2414C,2016MNRAS.455.1115H}, which decomposes phase-space into patches (`waterbags') of constant density, and by Liouville's theorem the dynamics can obtained from just evolving the boundary of these patches. The method required adaptive meshing with refinement for accurately capturing the stretching and folding that occurs in phase-space.

Here we make the case that IL can be a very useful numerical tool to study the Vlasov-Poisson system and gain insight into the evolution in the entire phase-space. The computational cost method is very cheap compared to other methods that resolve the entire phase-space, and also cheap compared to N-body/PM methods that use the same number of particles as IL's resolution elements. IL's Lagrangian, symplectic nature also helps resolve fine structures in phase-space evolution.

The paper is organized as follows.
In Section~\ref{sec:theory} we recall the basic theory of collisionless, self-gravitating systems, and lay out some notation.
In Section~\ref{sec:methods} we discuss the IL numerical method, and a technique to make it more memory efficient. Additionally, this section contains a description of a PM method, FV method and a moving-mesh (MM) method to which we compare the IL scheme. The MM scheme is our improvement to the FV scheme to reduce numerical diffusion, inspired by IL. 
In Section~\ref{sec:results} we present the results of simple numerical tests of Jeans instability and Landau damping, highlighting the strengths and limitations of each method.
In Section~\ref{sec:disc} we discuss the results, as well as the computational efficiency and costs of the method, and also outline extensions of the IL method for collisional systems.
Finally, we offer concluding remarks in Section~\ref{sec:conc}.

\section{Vlasov-Poisson equation}\label{sec:theory}
The dynamics of a collisionless, self-gravitating fluid is described by the well-known Vlasov-Poisson equation. The structure of the system is determined by the distribution function $f(\mathbf{r}, \mathbf{v}, t)$, which gives the phase-space density at each location $\mathbf{r}$, velocity $v$, and time $t$.
Physical density can be obtained from:
\begin{equation}
\rho(\mathbf{r},t) = \int (\mathbf{r}, \mathbf{v}, t) d^3\mathbf{v}.
\end{equation}
The gravitational potential, $\Phi(\mathbf{r},t)$, 
is given by Poisson's equation,
\begin{equation}
\nabla^2\Phi = 4\pi G\rho.
\end{equation}
The distribution function $f$ evolves via the collisionless Boltzmann equation,
\begin{equation}
\frac{\partial f}{\partial t}
+
\mathbf{v}\cdot\nabla f
-\nabla\Phi\cdot\frac{\partial f}{\partial\mathbf{v}} = 0
\end{equation}
which is statement of conservation of phase-space density.
Per Liouville's theorem, the phase-space distribution function is constant along the trajectories of the system.

\begin{figure}
\centering
\includegraphics[width=0.3\textwidth]{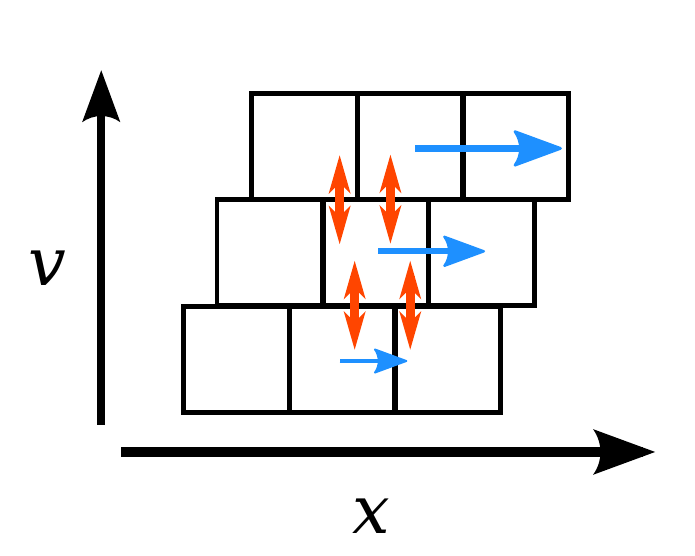}
\caption{A schematic representing the MM scheme. In each time-step, fluxes due to the gravitational acceleration (red arrows) are calculated across 4 interfaces per cell, while advection is treated exactly via mesh motion (blue arrows).}
\label{fig:mm}
\end{figure}

\section{Numerical Methods}\label{sec:methods}

Here we describe our IL method (Section~\ref{sec:IL}), 
and reference methods PM (Section~\ref{sec:PM}), FV (Section~\ref{sec:FV}), and MM (Section~\ref{sec:MM}) to which we compare it. MM is our moving mesh improvement upon the FV method, to improve the time-step criterion and amount of numerical diffusion.

\subsection{Integer Lattice (IL)}\label{sec:IL}

In the IL scheme, phase-space density distribution function $f$ is discretized onto a lattice. We describe the method in 1D physical space (2D phase-space), but it is readily extendible to any dimensions. The lattice nodes are separated a distance $\Delta x$ in position and $\Delta v$ in velocity.
Without loss of generality, choose units $\Delta x=\Delta v=1$.
Denote a lattice site with the coordinates $(x,v)$ (integers).
Each site has an associated value $f_{x,v}$ of the distribution function.
With each time step ($\Delta t$), the value $f_{x,v}$ associated with $(x,v)$
is updated to a new lattice site
via a `kick' step and a `drift' step:
\begin{equation}
v \leftarrow v + \lfloor\Delta t a\rceil
\label{eqn:x}
\end{equation}
\begin{equation}
x \leftarrow v + \lfloor \Delta t v \rceil
\label{eqn:v}
\end{equation}
where $\lfloor\cdot \rceil$ is the round-to-nearest integer operator.
The acceleration $a=\nabla\Phi$ can be obtained by any standard techniques to solve the Poisson equation, such as in Fourier space with a fast Fourier transform or with second-order finite difference discretization of the Laplacian operator. This step is cheap in the algorithm because the force has to be calculated at just the physical lattice points ($N^D$ sites as opposed to $N^{2D}$ sites in $D$ dimensions). At each physical lattice site $x$, the density $\rho_x$ is obtained by summing the contribution from each of the velocity sites at that location: $\rho_x = \sum_v f_{x,y}\Delta v$.
Figure~\ref{fig:algorithm} summarizes visually this simple IL algorithm.

As \cite{1995MNRAS.276..467S} point out, the update steps are simple first order, prior to rounding. But due to the rounding operation, using higher order methods is futile. 
The method has no round-off error, just truncation error due to the lattice resolution.
The integration scheme is completely reversible (i.e., unaffected by round-off).
Furthermore, due to the finite nature of the mesh, each particle will realize travelling on an orbit with some period, thus, the whole system has a long but finite period, realizing the Poincar\'e recurrence theorem.
The limit of a continuous system is approached as the resolution is increased.

It is important to point out a difference between our method and that of \cite{1995MNRAS.276..467S}.
\cite{1995MNRAS.276..467S} fix the time-step to 
$\Delta t = \Delta x / \Delta v$, 
so that advection at the lattice site $v = \Delta v$ is resolved (i.e., $\Delta v$ is the minimum velocity that can be advected on the lattice by the scheme, due to rounding).
Therefore, the equivalent of Equation~\ref{eqn:v} of \cite{1995MNRAS.276..467S} does not use a $\lfloor\cdot\rceil$ operator.
However, we find that this choice of time-step is not general, and too restrictive in certain cases to be able to resolve the dynamical timescale given by $T= v / a$, where $v$ and $a$ are the characteristic velocities and accelerations of the system.
That is, to accurately resolve a system, one needs to pick $\Delta t < T$.
In a general simulation with a predetermined choice of $\Delta x$ and $\Delta v$, it may be the case that $T < \Delta x / \Delta v$ (unless one reduces the $\Delta x$ spacing or increases the $\Delta v$ spacing, which may not be desirable).
This necessitates that one pick a smaller time-step than $\Delta x / \Delta v$, 
which introduces the need for the 
$\lfloor\cdot\rceil$ operator in Equation~\ref{eqn:v}.
In the case $\Delta t< \Delta x / \Delta v$, the smallest velocity $v_{\rm min}$ advected by the scheme will be $v_{\rm min} > \Delta v$. In general, $v_{\rm min} = n\Delta v$, $n$ an integer. Hence this affects the staggered pattern of the `kick' step in the algorithm
(see Figure~\ref{fig:algorithm}, which has $n=2$). One must choose $1 \leq n \ll N_x$ to adequately resolve the shearing motion due to advection in each time-step. 

In other words, if one restricts themselves to use a time-step $\Delta t = \Delta x / \Delta v$, this is not general enough to ensure the dynamical timescale is resolved and one then needs to either reduce the $\Delta x$ spacing (i.e., increase spatial resolution $N_x$) or increase the $\Delta v$ spacing (i.e., decrease velocity resolution $N_v$) to decrease the time-step $\Delta t$ in order to obtain an accurate simulation. The formulation we present in this manuscript is general for arbitrary grid spacing. The frequency at which the original time-step $\Delta x / \Delta v$ is larger than the dynamical timescale can be anywhere between $100$~per~cent (which is the case for the simulations presented in Section~\ref{sec:results}) and $0$~per~cent, depending on what values one chooses for the resolution: $N_x$ and $N_v$.

Note, in general, unlike most other numerical methods, the time-step cannot be arbitrarily small, otherwise no advection is resolved. In general, it is best to ensure $v_{\rm min} \ll v_{\rm max}$, where $v_{\rm max}$ is the maximum velocity lattice point.

It is worth stressing, as pointed out in \cite{1995MNRAS.276..467S}, that the
integer lattice method adheres to important theoretical properties of general lattice approaches described in \cite{1992PhyD...56....1E}.  Namely, the rounding operations in Equations~\ref{eqn:x} and \ref{eqn:v} produce a map on the lattice that is one-to-one, and the map is itself Hamiltonian. Furthermore, the lattice map approaches the original (continuous) map as the lattice spacing is decreased. Finally, the method has no floating-point round-off errors due to integer arithmetic, and can be reversed exactly.

\subsubsection{Improving the memory efficiency of IL}

Ordinarily, the memory requirement for IL is $\mathcal{O}(N_x^DN_v^D)$, where $N_x$ is the spatial resolution per linear dimension, $N_v$ is the velocity resolution per linear dimension, and $D$ is the physical dimension of the problem.
That is, the distribution function $f_{x,v}$ is stored at each lattice site.
This intense memory requirement is the bane of all 6D phase-space resolving methods.
However, it is possible to significantly reduce the memory requirement in the case of IL.
At any time, $f_{x,v}$ can simply be recovered by tracing any lattice site $(x,v)$ back to its initial location (at which $f$ can be computed via a function $f_0(x,v)$ specifying the initial condition) by undoing the `drift' and `kick' operations. Note, this is possible because of the exact reversible nature of the method, unaffected by round-off errors.
This reversal can be done efficiently by storing the $N_x^D$ acceleration integers at each time-step.
The computational cost is increased a bit:
instead of an $\mathcal{O}(1)$ memory lookup of $f_{x,v}$, we require, at $N_t$ time-steps, $\mathcal{O}(N_t)$  lookups and operations to undo the `drifts' and `kicks'.
However, there can be a drastic gain in memory requirements.
One must store $\mathcal{O}(N_x^DN_t)$ accelerations rather than
$\mathcal{O}(N_x^DN_v^D)$ lattice sites.
Many applications have $N_x\sim N_v\sim N_t\sim N$ (in fact, our simulations have $N_t\ll N_x$).
In this case, the memory requirement is reduced from $\mathcal{O}(N^6)$
to  $\mathcal{O}(N^4)$ for a full 3D physical system. 
The added computational cost is a small price to pay for the reduction of memory usage, especially in the current age of supercomputing, where computational power is cheap but memory is expensive.

The above approach seems limited if the number of time-steps required in the simulation is large $N_t > N_v^3$. However, there can be workarounds to this limitation as well.
After $N_t$ steps (if $N_t$ is large), it may be beneficial to compute the full phase-space and use a compression algorithm with fast lookup capabilities to store it in a memory efficient way. Then, 
the above technique of storing accelerations can be repeated for another round of time-steps, but for deducing $f_{x,v}$ one now only need to undo the `drifts' and `kicks' to this new point in time where the full phase-space state is stored. The details of such a method is beyond the scope of the present paper and are left for future work.

We note that here we assumed that the initial conditions $f_0(x,v)$ are simple in that they can be described by evaluating a function, or require a small amount of memory to be loaded. If the initial conditions are complex and need a large amount of memory to be 
loaded, then they need to be approximated using a small number of interpolating functions or using compression algorithms with fast lookup capabilities, otherwise the implementation strategies to improve memory-efficiency described here offer no advantage.

\subsection{Particle Mesh (PM)}\label{sec:PM}

We compare the IL scheme to a basic PM scheme described here.
The PM scheme consists of $N$ particles updated via a $2^{{\rm nd}}$-order symplectic leap-frog scheme (`kick'-`drift'-`kick'):
\begin{equation}
v \leftarrow v + 0.5\Delta t a
\end{equation}
\begin{equation}
x \leftarrow x + \Delta t v
\end{equation}
\begin{equation}
v \leftarrow v + 0.5\Delta t a
\end{equation}
To calculate the acceleration $a$, we use a cloud-in-cell (CIC) approach. 
Physical space is discretized into $M$ grid points. We choose $N=M^2$.
Each particle contributes to the density at each physical grid point via standard CIC weighting. The Poisson equation is solved on the physical grid points as in Section~\ref{sec:IL}. The acceleration at each particle may be computed by interpolating the acceleration known at the physical grid points to the particle locations via $3^{\rm rd}$-order cubic spline interpolation.

\begin{figure*}
\centering
\begin{tabular}{cccc}
 & \large{$t=3$} & \large{$t=9$} & \multirow{5}{*}{\includegraphics[height=0.914\textheight]{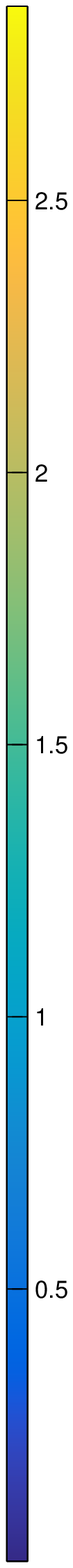}} \\
\rotatebox{90}{{\large\qquad\qquad\qquad\quad IL}}\,\,\, &
\includegraphics[width=0.4\textwidth]{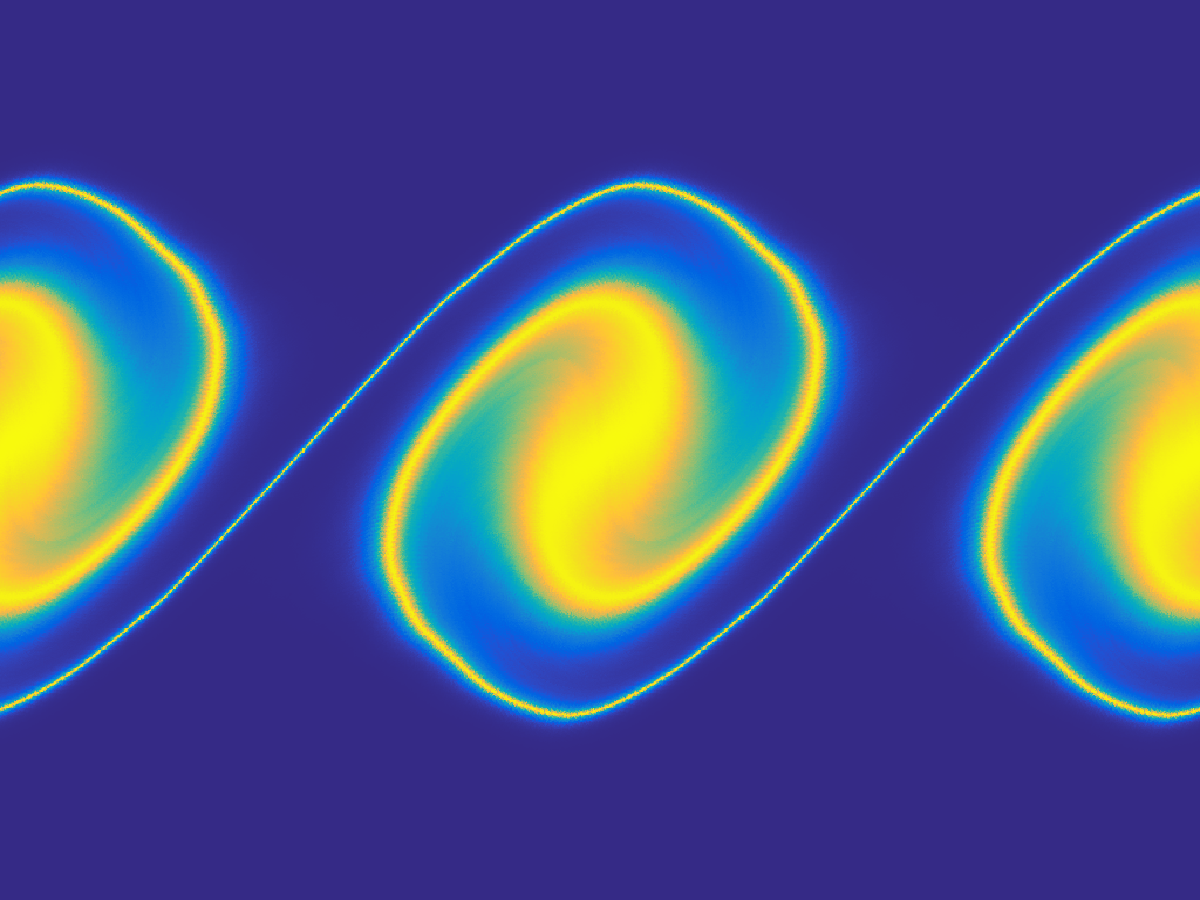} &
\includegraphics[width=0.4\textwidth]{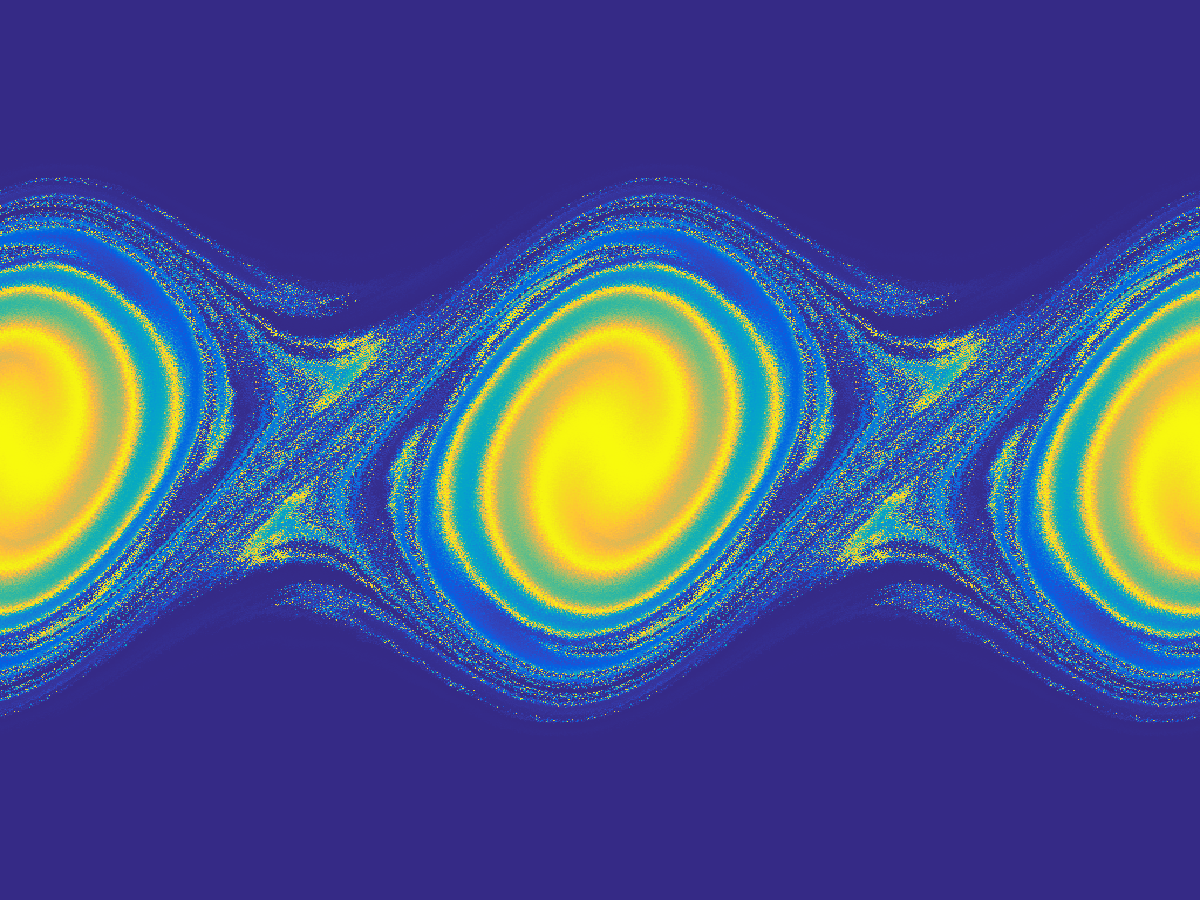} \\
\rotatebox{90}{{\large\qquad\qquad\qquad\quad PM}}\,\,\, &
\includegraphics[width=0.4\textwidth]{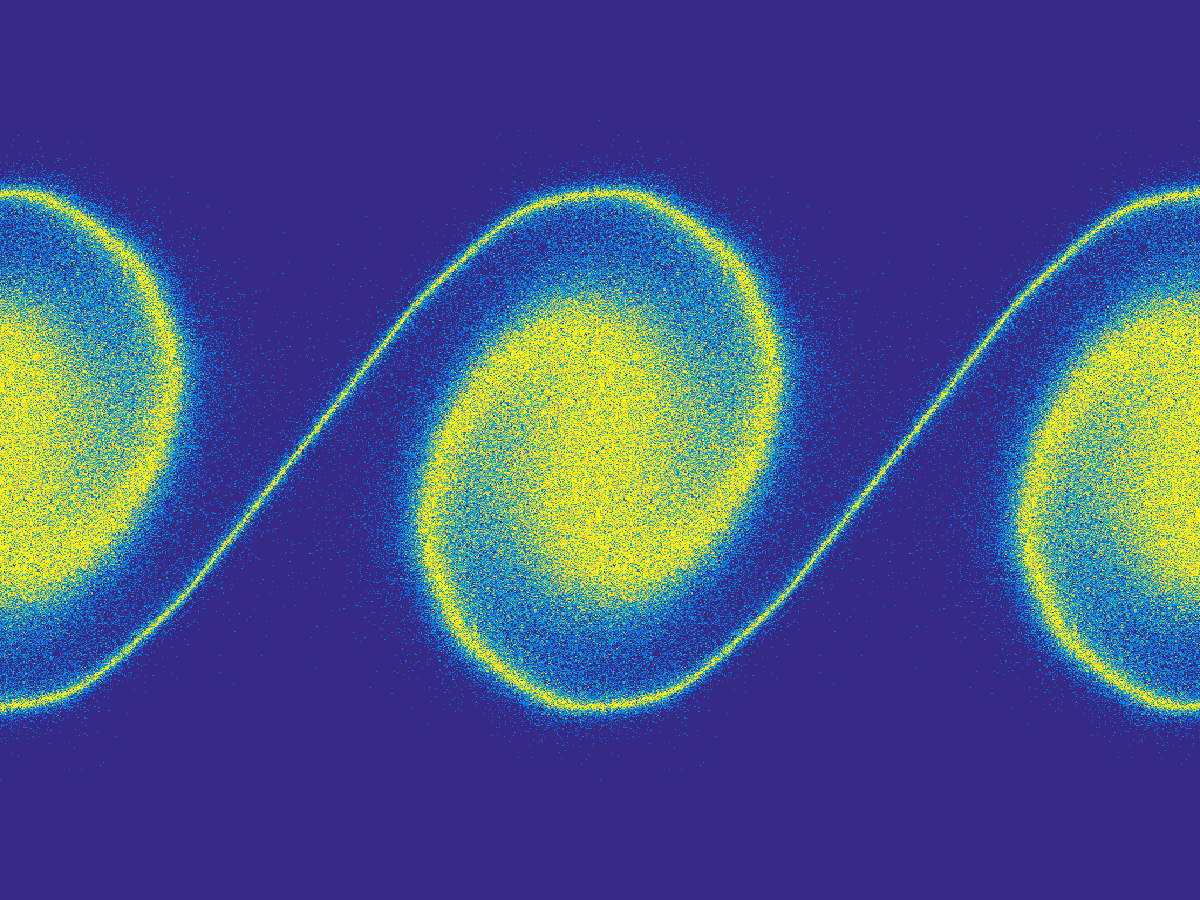} &
\includegraphics[width=0.4\textwidth]{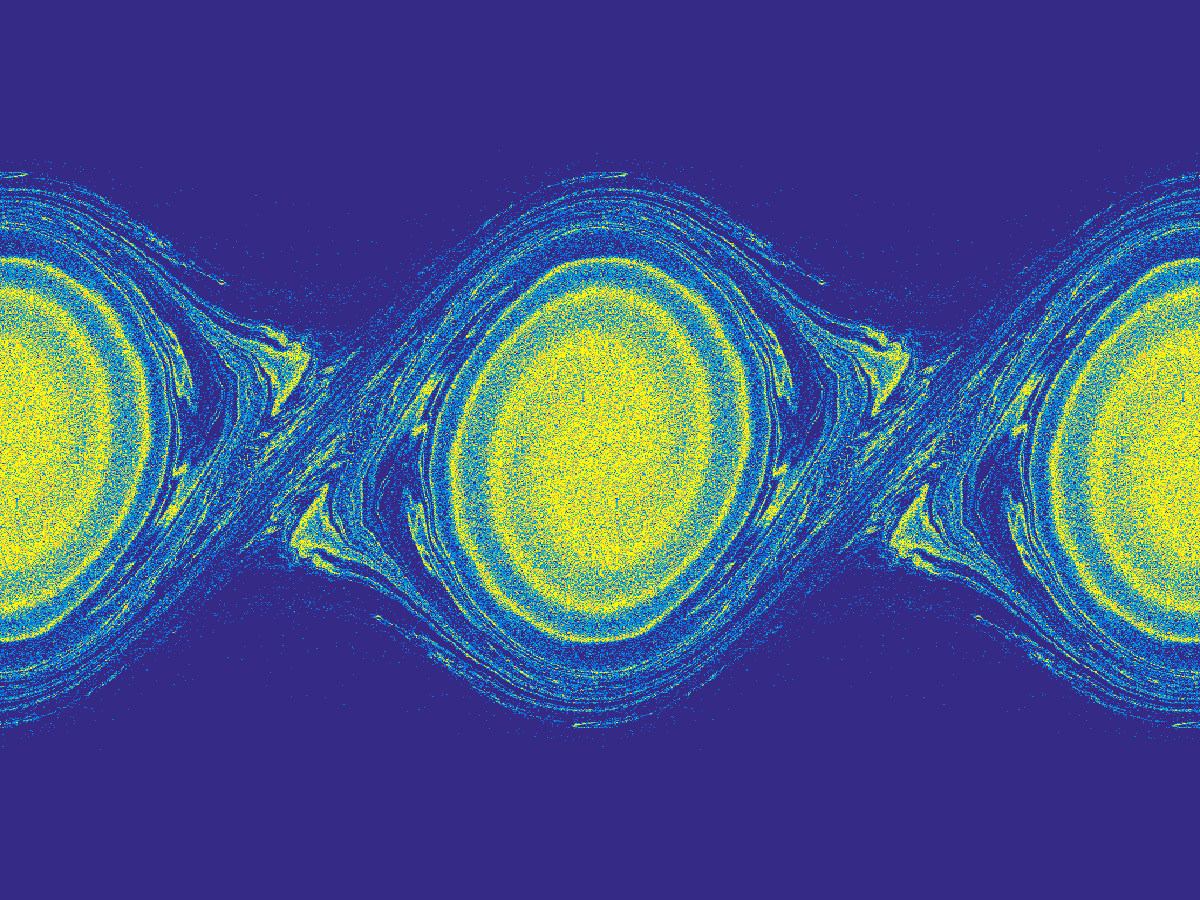} \\
\rotatebox{90}{{\large\qquad\qquad\qquad\quad FV}}\,\,\, &
\includegraphics[width=0.4\textwidth]{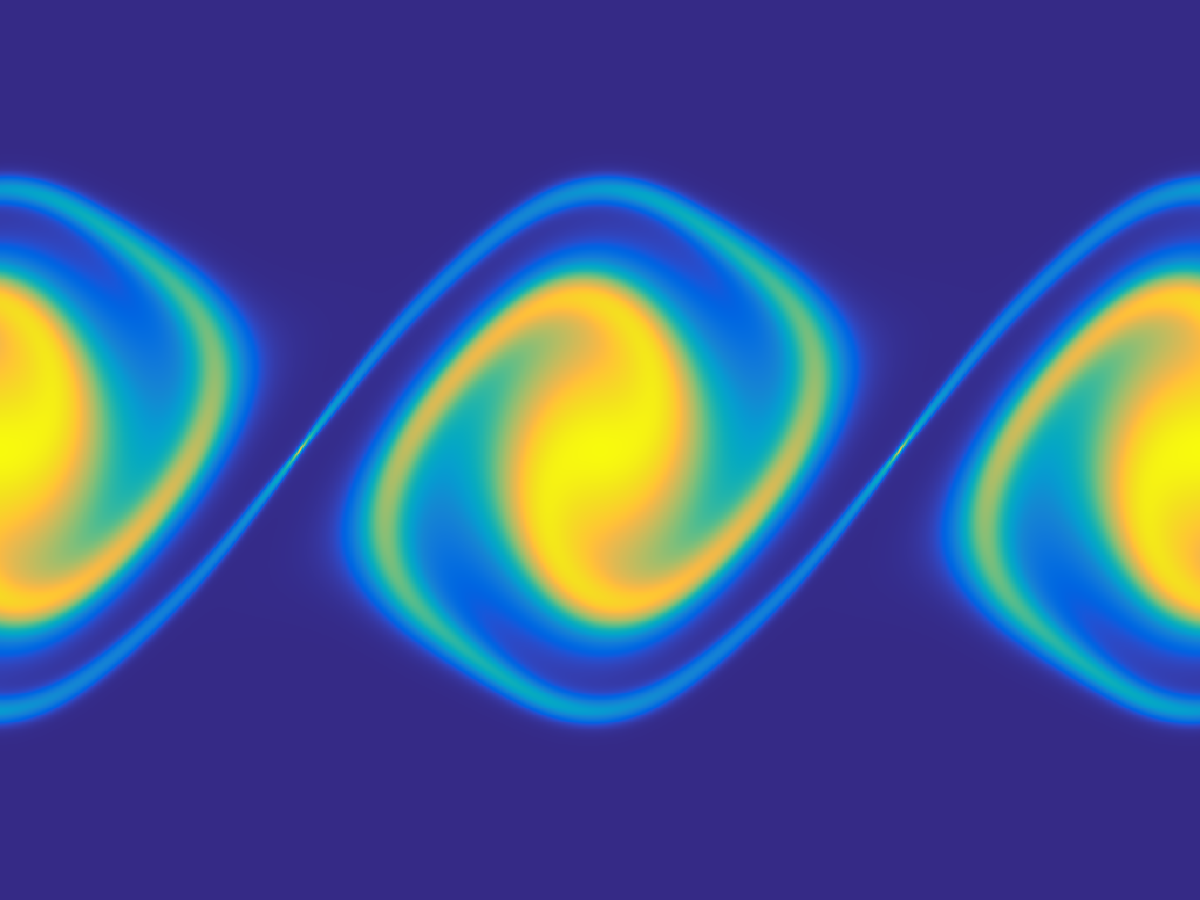} &
\includegraphics[width=0.4\textwidth]{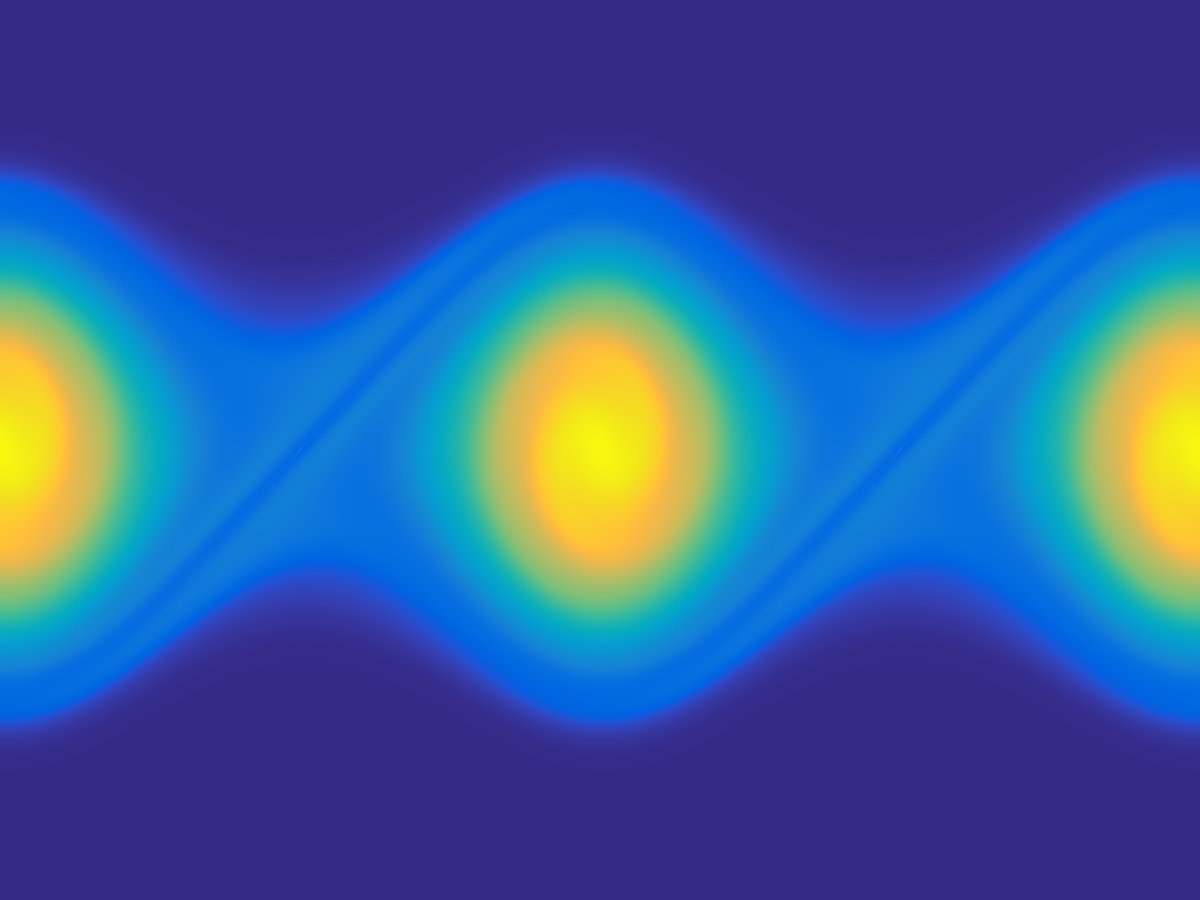} \\
\rotatebox{90}{{\large\qquad\qquad\qquad\quad MM}}\,\,\, &
\includegraphics[width=0.4\textwidth]{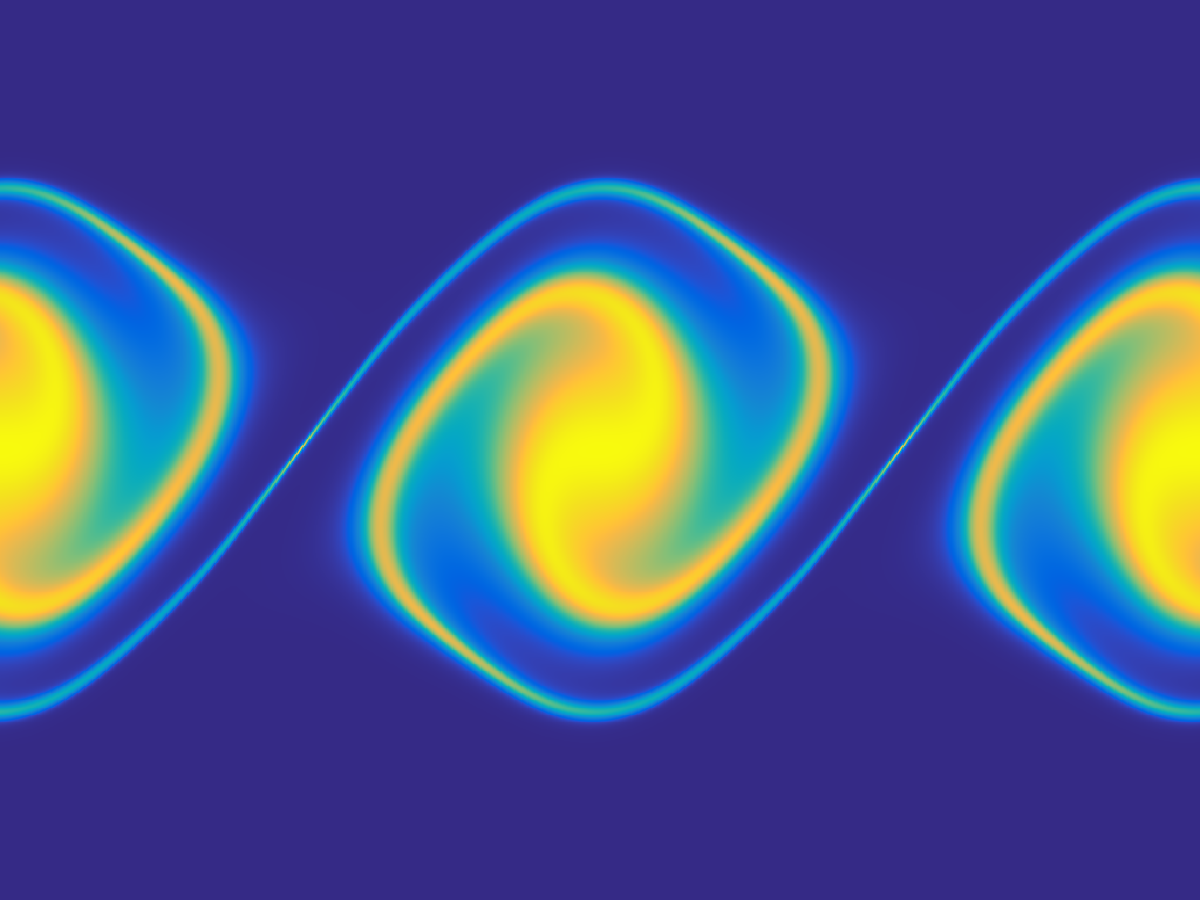} &
\includegraphics[width=0.4\textwidth]{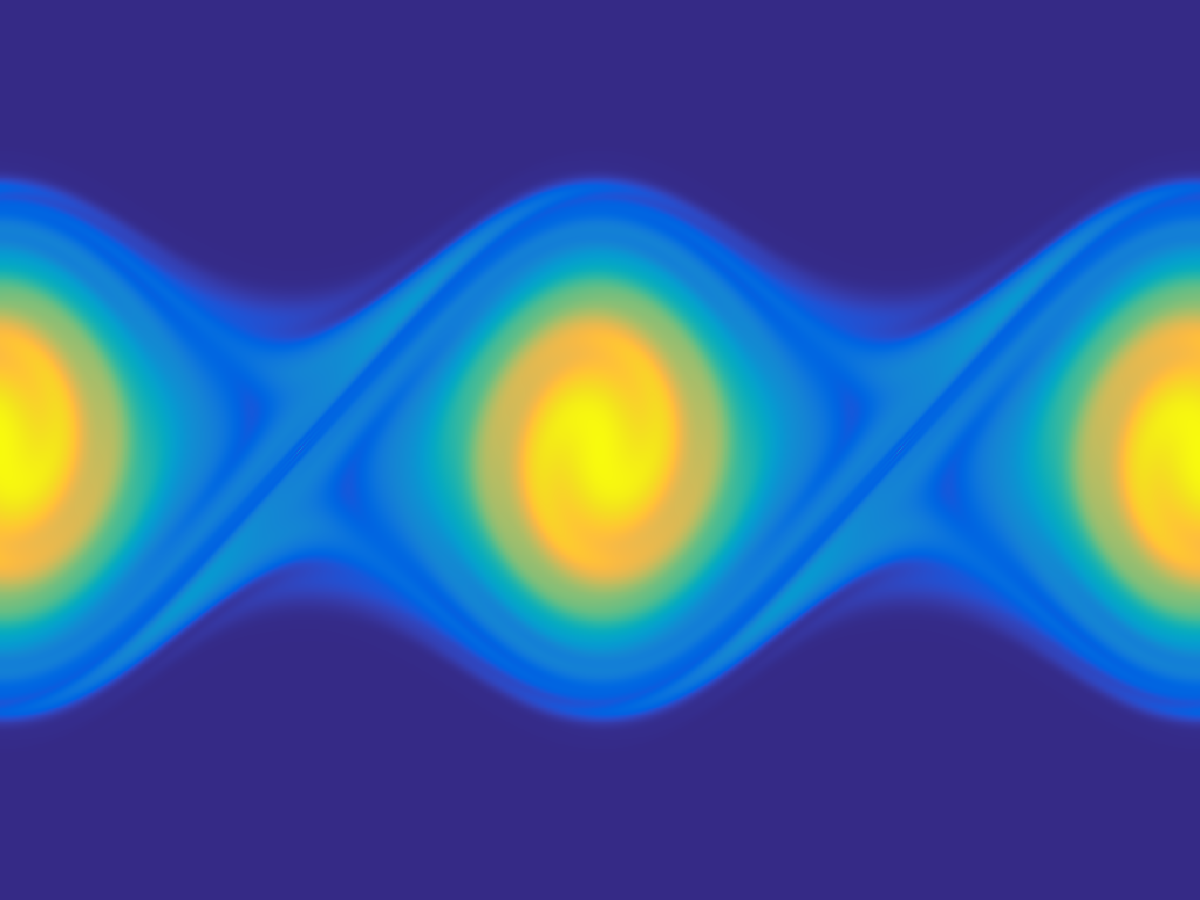} \\
\end{tabular}
\caption{
Simulated phase-space of the Jeans instability problem (section~\ref{sec:jeans}) with the various methods at times $t=3,9$ using a resolution of $N_x=N_v=1024$ (equivalently, $N=1024^2$ particles for PM).
}
\label{fig:jeans}
\end{figure*}

\begin{figure*}
\centering
\begin{tabular}{cccc}
 & \large{$t=5$} & \large{$t=25$} & \multirow{5}{*}{\includegraphics[height=0.914\textheight]{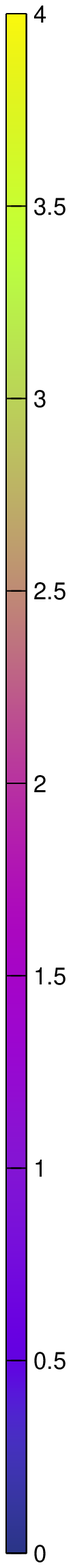}} \\
\rotatebox{90}{{\large\qquad\qquad\qquad\quad IL}}\,\,\, &
\includegraphics[width=0.4\textwidth]{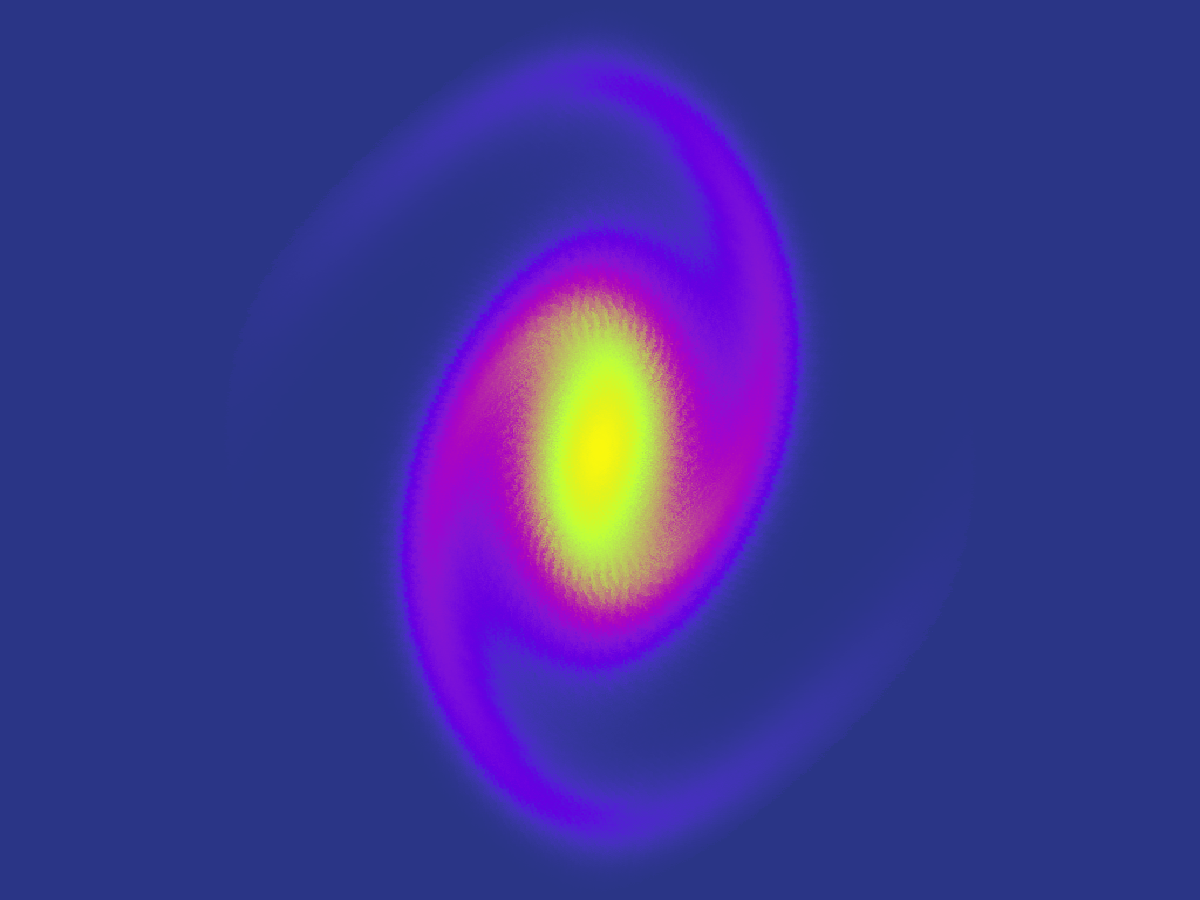} &
\includegraphics[width=0.4\textwidth]{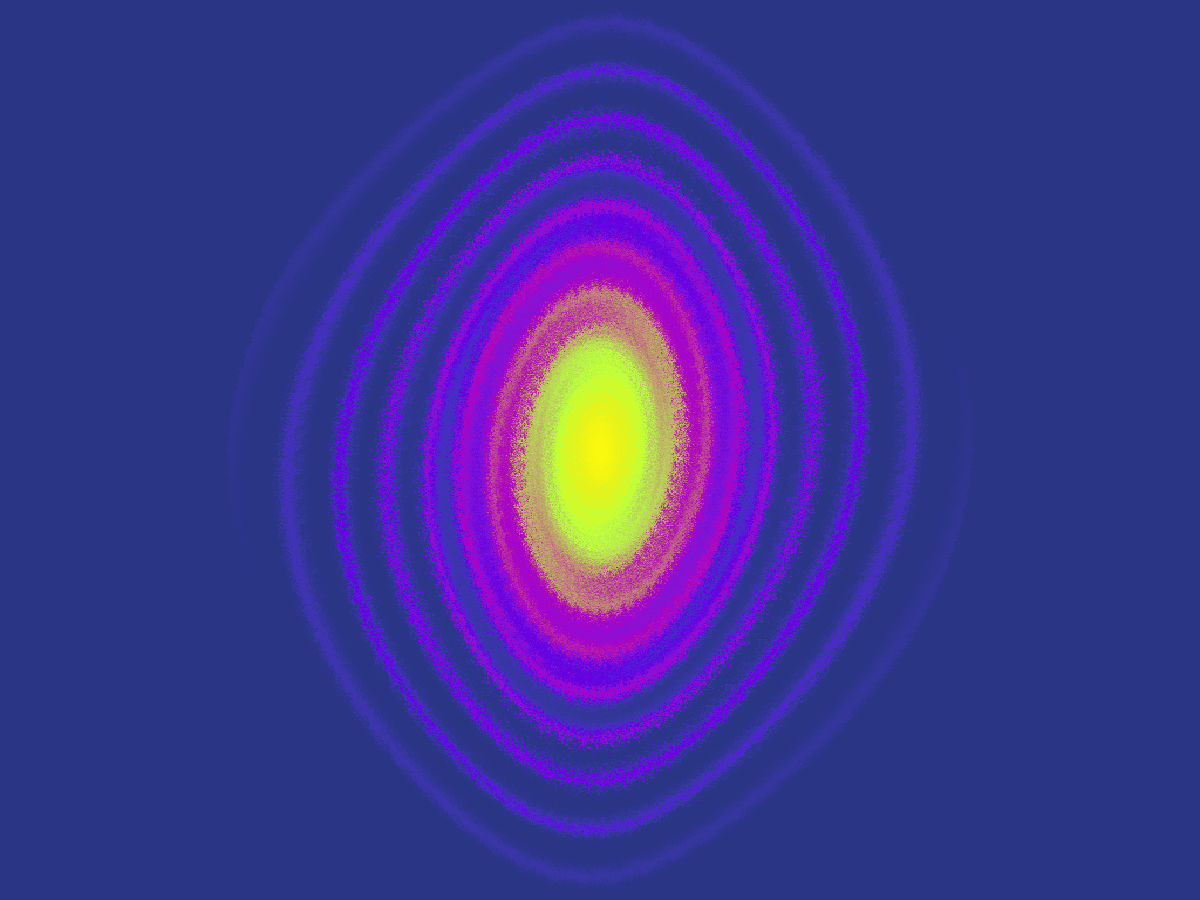} \\
\rotatebox{90}{{\large\qquad\qquad\qquad\quad PM}}\,\,\, &
\includegraphics[width=0.4\textwidth]{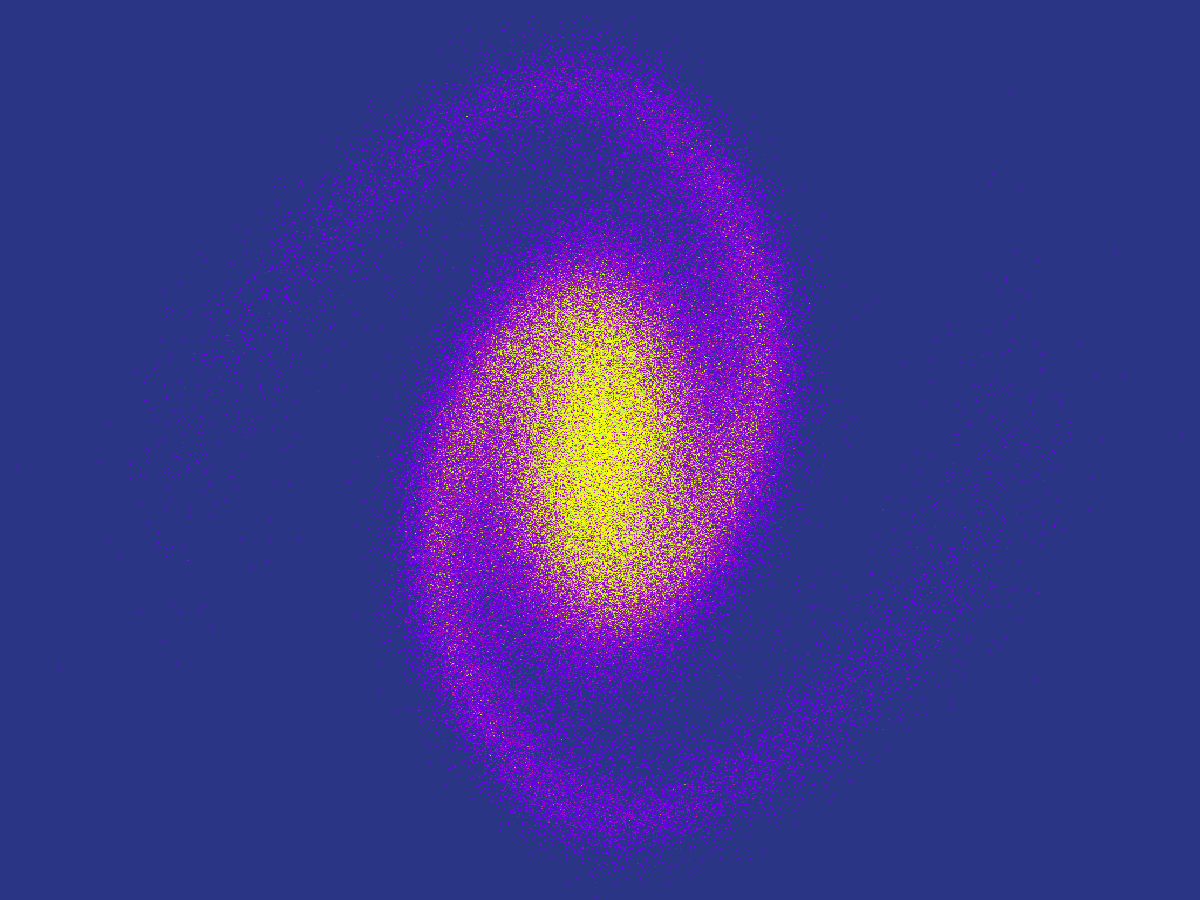} &
\includegraphics[width=0.4\textwidth]{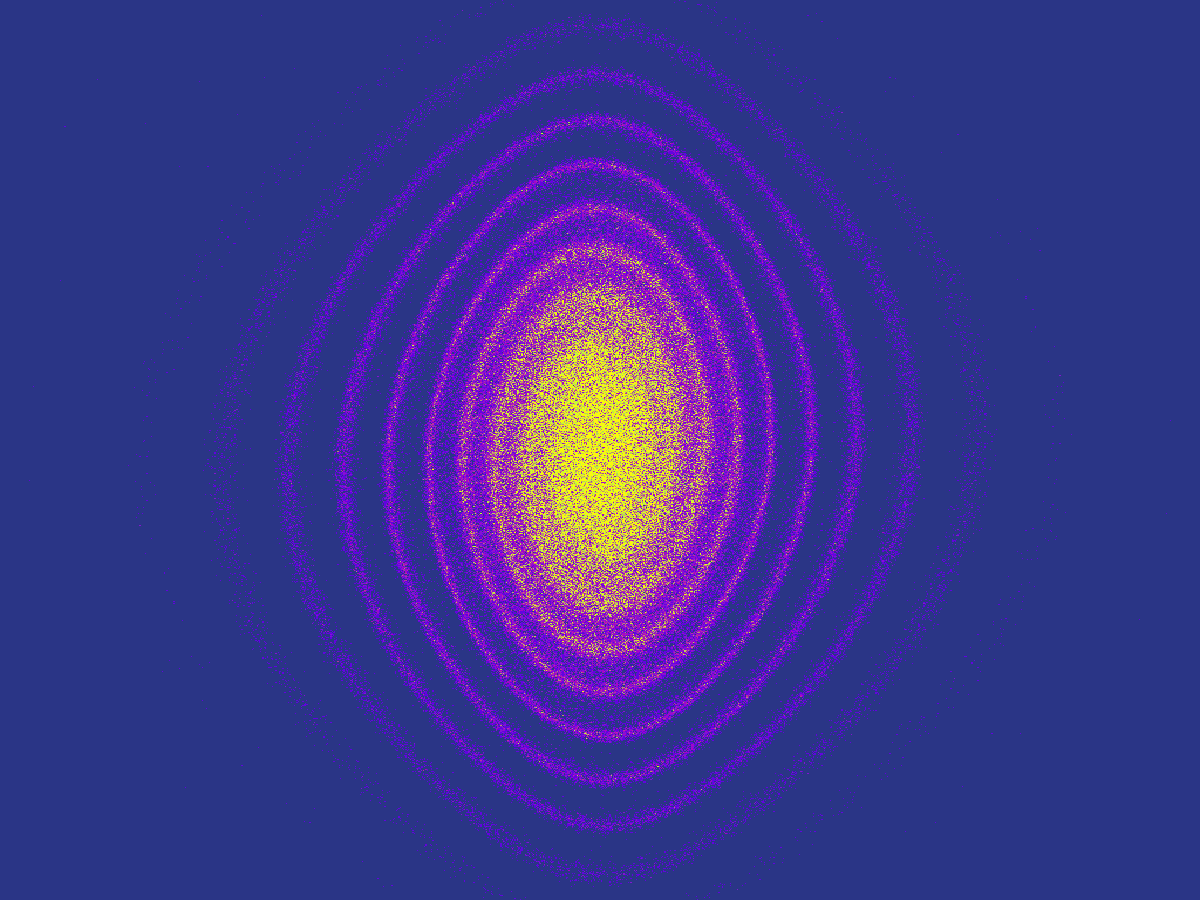} \\
\rotatebox{90}{{\large\qquad\qquad\qquad\quad FV}}\,\,\, &
\includegraphics[width=0.4\textwidth]{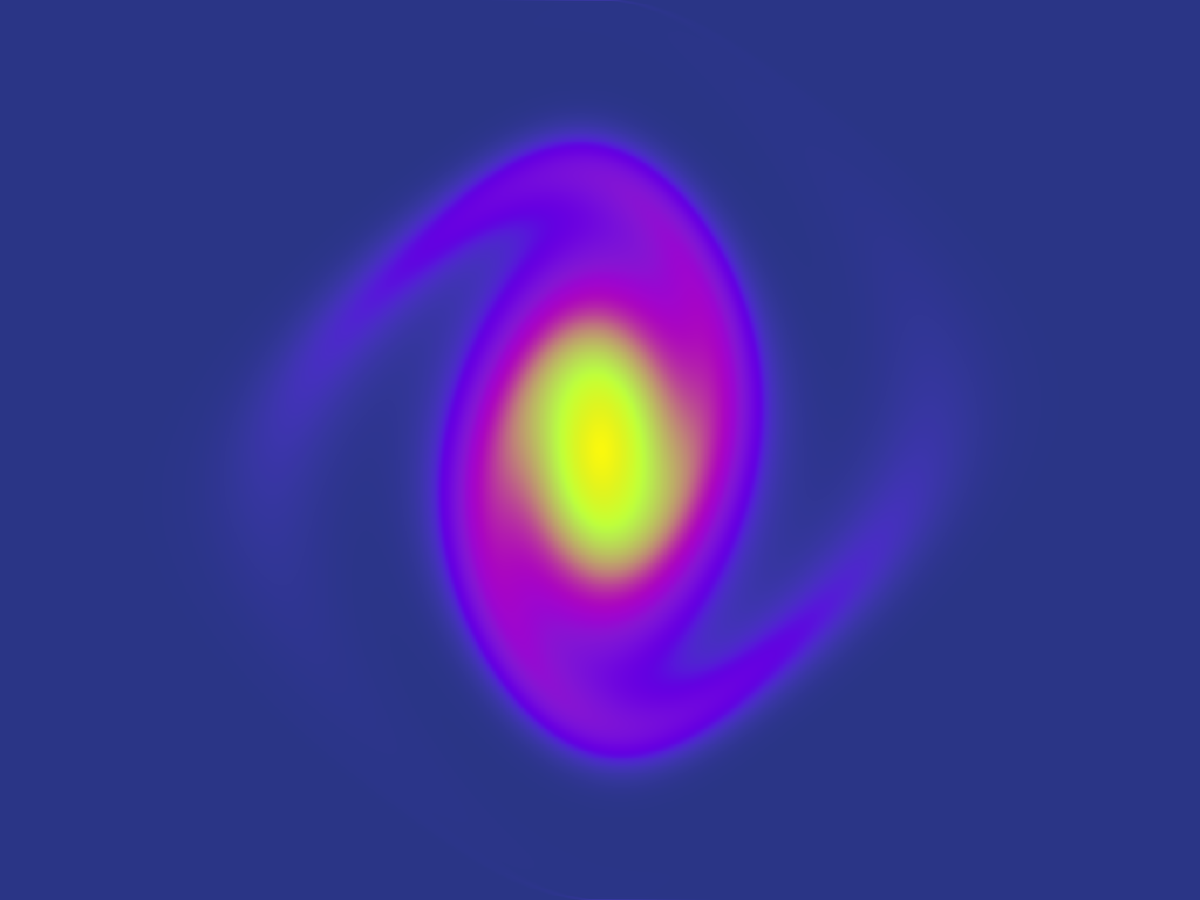} &
\includegraphics[width=0.4\textwidth]{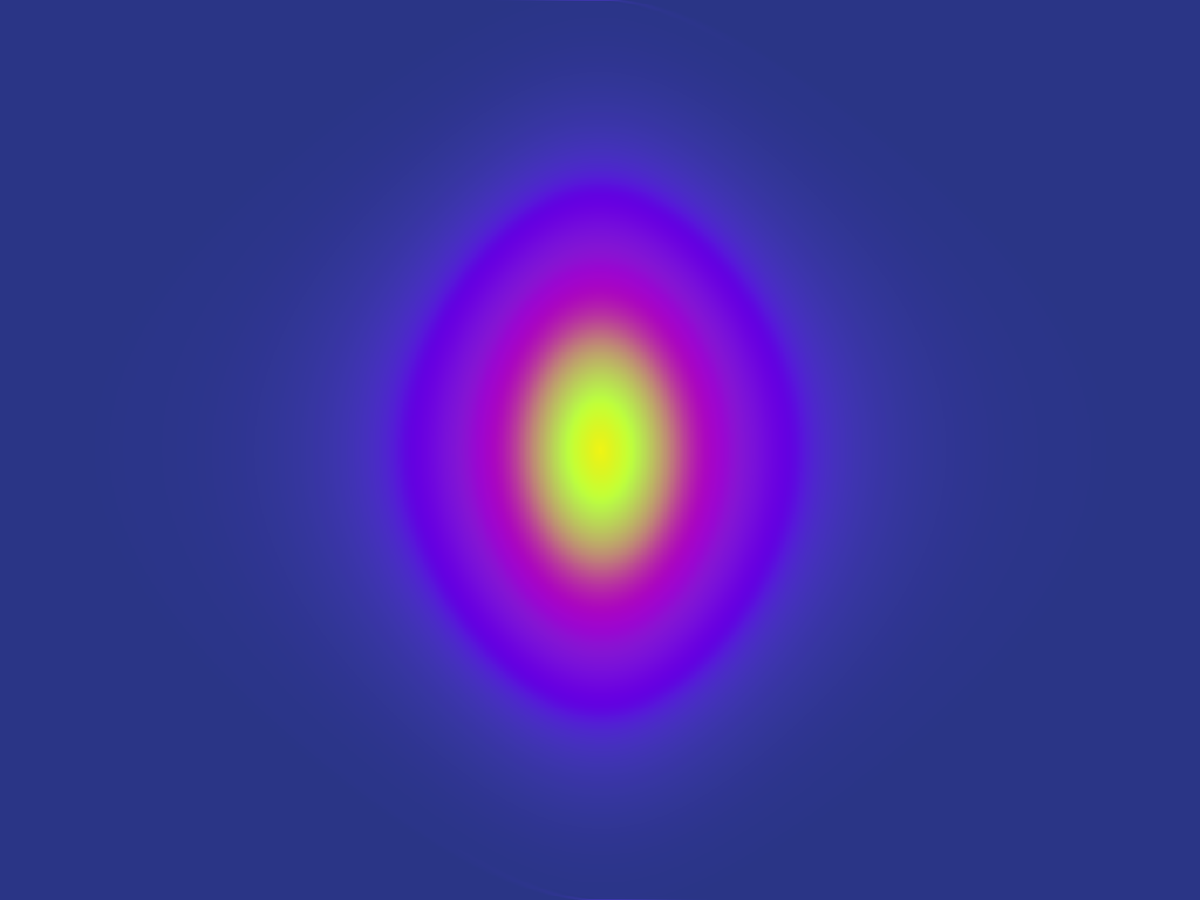} \\
\rotatebox{90}{{\large\qquad\qquad\qquad\quad MM}}\,\,\, &
\includegraphics[width=0.4\textwidth]{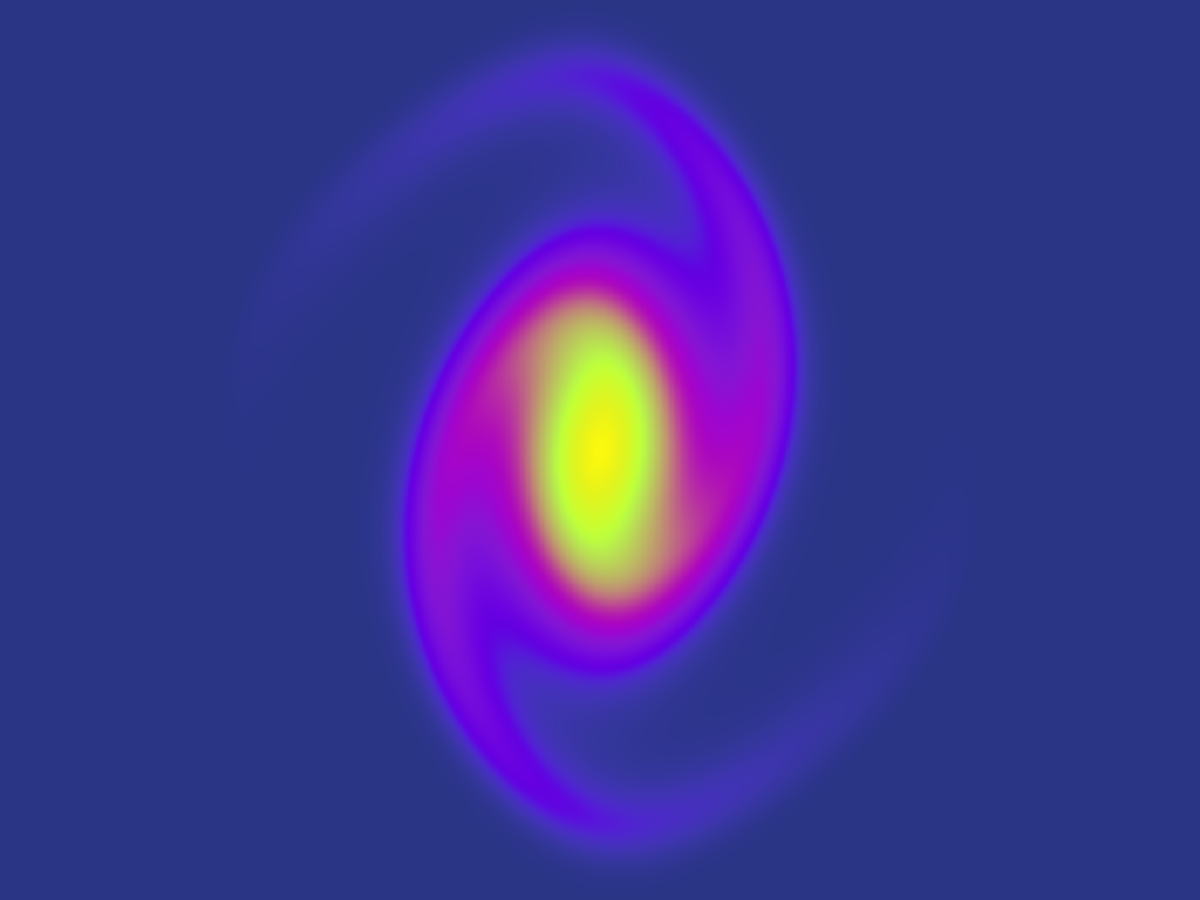} &
\includegraphics[width=0.4\textwidth]{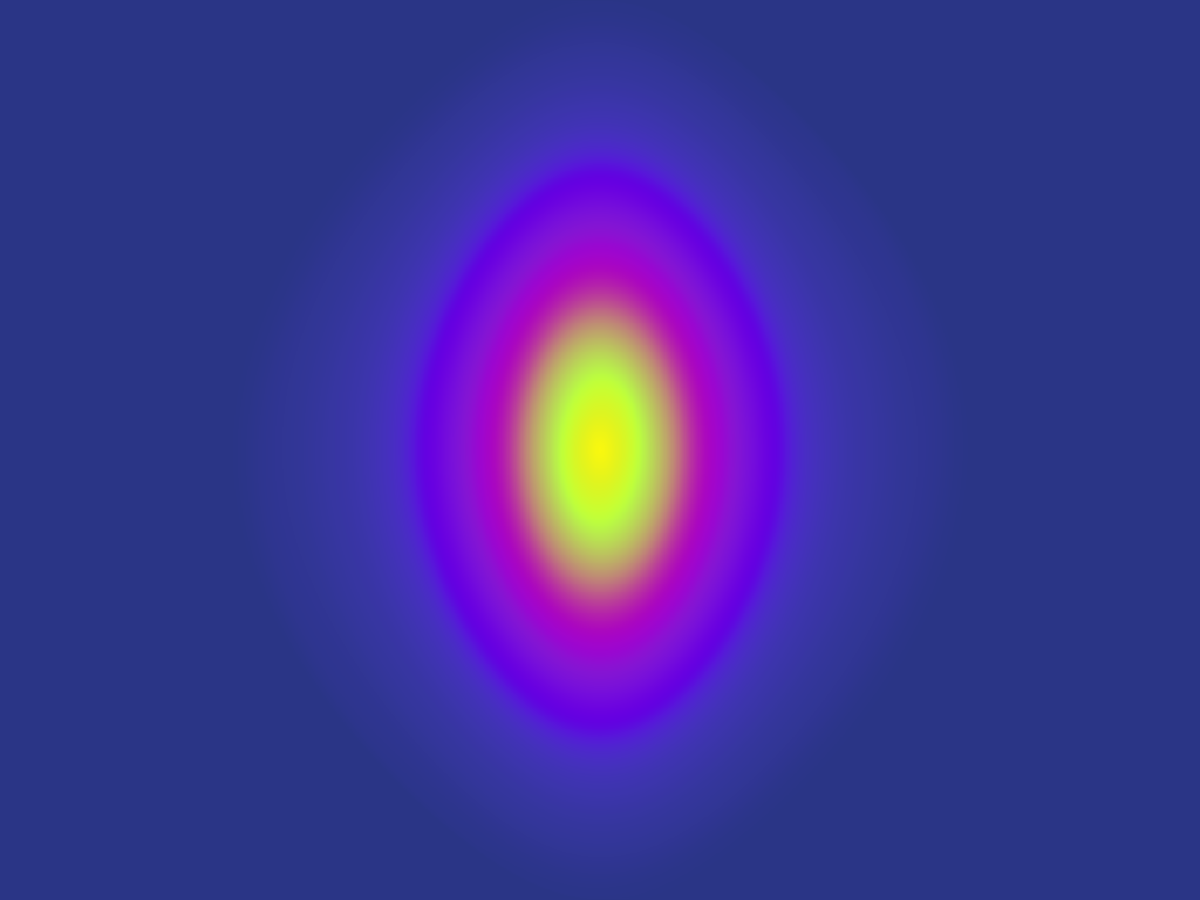} \\
\end{tabular}
\caption{
Simulated phase-space of the initial Gaussian problem (section~\ref{sec:landau}) with the various methods at times $t=5,25$ using a resolution of $N_x=N_v=1024$ (equivalently, $N=1024^2$ particles for PM).
}
\label{fig:landau}
\end{figure*}

\subsection{Finite Volume (FV) and Moving Mesh (MM)}\label{sec:FV}

We also construct a simple $2^{{\rm nd}}$-order FV scheme as a reference scheme to which we can compare IL. Here we'll briefly outline the basic steps of this common approach for solving the fluid equations. For details and improvements to the scheme, see \cite{2013ApJ...762..116Y}.
First, we discretize phase-space into cells of dimensions $\Delta x$ by $\Delta v$.
The cell-averaged distribution function at each site, $f_{x,v}$, is updated by computing fluxes across the faces neighbouring cells (i.e., material is transferred between cells, making the method conservative).
The cells are updated in a second-order time-symmetric fashion (analogous to `kick'-`drift'-`kick'):
\begin{equation}
v \leftarrow v - 0.5 \frac{\Delta t}{\Delta v} \sum_{n\in{\rm neighbours}} \mathcal{F}_{v,n}
\end{equation}
\begin{equation}
x \leftarrow x -  \frac{\Delta t}{\Delta x} \sum_{n\in{\rm neighbours}} \mathcal{F}_{x,n}
\end{equation}
\begin{equation}
v \leftarrow v - 0.5 \frac{\Delta t}{\Delta v} \sum_{n\in{\rm neighbours}} \mathcal{F}_{v,n}
\end{equation}
The $\mathcal{F}_{v,n}$ and $\mathcal{F}_{x,n}$ are numerical fluxes, which we compute via the local Lax-Friedrichs flux across a face determined by a `left' (L) face state and a `right' (R) face state. 
The fluxes are given by 
\begin{equation}
\mathcal{F}_{v} = \frac{1}{2}\left(f_{\rm L}a_{\rm L} + f_{\rm R}a_{\rm R}\right)
-\frac{{\rm min}(|a_{\rm L}|,|a_{\rm R}|)}{2}\left(f_{\rm R} - f_{\rm L}\right)
\end{equation}
\begin{equation}
\mathcal{F}_{x} = \frac{1}{2}\left(f_{\rm L}v_{\rm L} + f_{\rm R}v_{\rm R}\right)
-\frac{{\rm min}(|v_{\rm L}|,|v_{\rm R}|)}{2}\left(f_{\rm R} - f_{\rm L}\right)
\end{equation}
These numerical fluxes are a combination of averaged L and R state fluxes plus a diffusion term for numerical stability.
The L and R states are obtained from linearly interpolating $f_{x,v}$, $a_{x,v}$, $v_{x,v}$ from cell centers to face center-of-masses from either side of the cells. 
The gradients required for linear interpolation are computed via $2^{{\rm nd}}$-order central finite difference stencils.

This method is an Eulerian scheme as opposed to Lagrangian. 
This makes the time-steps more restrictive.
For stability, we need to satisfy the CFL condition:
\begin{equation}
\Delta t = C{\rm min}\left( \frac{\Delta x}{v_{\rm max}}, \frac{\Delta v}{a_{\rm max}} \right)
\end{equation}
where $C\leq 1$ is a parameter to control the time-step, which we set to $C=0.5$.

\subsubsection{MM improvement of FV}\label{sec:MM}

Here we describe the MM method, a moving mesh improvement of FV which makes the scheme less diffusive and also loosens the time-step criterion to:
\begin{equation}
\Delta t = C{\rm min}\left( \frac{\Delta x}{\Delta v}, \frac{\Delta v}{a_{\rm max}} \right)
\end{equation}
In our particular moving-mesh approach, we will resolve advection exactly with mesh motion. This makes the method semi-Lagrangian and reduces the need to compute the flux $\mathcal{F}_{x}$. In practice we find this scheme is a lot less diffusive compared to the base FV scheme. Additionally, the scheme is now Galilean-invariant, in the sense that the same solution is obtained from boosting the initial conditions by a constant velocity and the amount of numerical diffusion and time-step criterion is unchanged. The Eulerian FV scheme does not satisfy this property: velocity boosting the initial conditions reduces the CFL-stable time-step criterion and may affect the code's ability to resolve instabilities due to truncation error \citep{2010MNRAS.401..791S}.

Simply, the calculation of the flux $\mathcal{F}_{x}$ in the `drift' step is replaced by moving each mesh cell at velocity $v$. 
This makes the phase-space cells simply shear past each other, making it easy to compute the mesh structure and identify neighboring cells at any given time in the simulation.
See Figure~\ref{fig:mm} for an illustration of the drifting motion.
The calculation of the fluxes $\mathcal{F}_{v}$ in the `kick' step
needs to be modified slightly, but straightforwardly. At a general time, each cell now shares partial faces with $4$ other cells instead of $2$. 
This requires computation of $4$ fluxes. For second-order accuracy, one interpolates the L and R states to the center-of-masses of the partial faces to feed into the numerical flux solver. 

We point out that we could adopt more complicated and arbitrary mesh motion in general. Another suitable choice would be a moving Voronoi mesh, which deforms continuously in a quasi-Lagrangian fashion, as has been used to solve the Euler equations \citep{2010MNRAS.401..791S,2011ApJS..197...15D}.
Such a choice would improve the time-step criterion further and reduce diffusion.
Constructing a Voronoi mesh in 6D would be costly but possible. However, the efficiency of such a method could be further improved with meshless approximations to Voronoi meshes \citep{2011MNRAS.414..129G,2015MNRAS.450...53H}.

\section{Results}\label{sec:results}

We present two numerical tests (Section~\ref{sec:jeans} and Section~\ref{sec:landau}) to demonstrate that IL can successfully be used to simulate dynamics and reveal fine-grained structure in phase-space, and highlight the advantages and disadvantages of the IL, PM, FV, and MM methods.

\subsection{Jeans Instability}\label{sec:jeans}

In this test, from \cite{2013ApJ...762..116Y}, we study a one-dimensional, infinite, self-gravitating system.
The domain is:
\begin{equation}
\begin{cases}
-L/2\leq x \leq L/2, \\
-V\leq v \leq V
\end{cases}
\end{equation}
$x$ is periodic on a domain of size $L$, and the maximum velocity is $V=L/T$, 
where $T=(G\overline{\rho}^{-1/2})$ is the dynamical time, and $\overline{\rho}$ is the average density of the system. Without loss of generality, we use units of $L=V=T=1$.
The initial distribution function is:
\begin{equation}
f(x,v,t=0) = \frac{\overline{\rho}}{(2\pi\sigma^2)^{1/2}}
{\rm exp}\left(-\frac{v^2}{2\sigma^2}\right)
\left(1 + A \cos(kx)\right)
\end{equation}
where $\sigma$ is the velocity dispersion, $A$ is the amplitude of the density fluctuation, and $k$ is the wave number of the density fluctuation.
If the wave number $k$ is smaller than the critical Jeans wavenumber $k_{\rm J}$, which is given by:
\begin{equation}
k_{\rm J} = \left(\frac{4\pi G\overline{\rho}}{\sigma^2}\right)^{1/2},
\end{equation}
then the Jeans instability is triggered. Else, the density fluctuation damps by Landau damping, analogous to the process from plasma physics.

Due to the periodic boundary conditions, $k=nk_0$, $k_0=2\pi/L$, $n$ an integer. We choose $n=2$.
We simulate the Jeans instability triggered by $k/k_{\rm J}  = 0.5$. The amplitude of the perturbation is set to $A=0.01$.

Our IL, FV, and MM simulations use a resolution of $N_x=N_v=1024$. The PM simulation uses $N=1024^2$ particles, equivalent to the number of cells in the other simulations. 

Figure~\ref{fig:jeans} shows the simulated distribution function at $t=3$ and $t=9$, showing the non-linear regime of the instability.

\subsection{Gaussian: Landau Damping}\label{sec:landau}

Here we simulate an initially Gaussian distribution of matter in phase-space near thermal equilibrium, a numerical test from \cite{2005MNRAS.359..123A,2014MNRAS.441.2414C}.
The initial condition is given by
\begin{equation}
f(x,v,t=0) = 4{\rm exp}\left(-(x^2+v^2)/0.08)\right)
\end{equation}
in the domain
\begin{equation}
\begin{cases}
-1\leq x \leq 1, \\
-1\leq v \leq 1
\end{cases}
\end{equation}
This system converges smoothly to a
quasi-steady-state via Landau damping. 
The system develops very thin `arms' in phase-space that spiral and wind up.
The fine structure in the solution makes it a useful test problem.

Again, our IL, FV, and MM simulations use a resolution of $N_x=N_v=1024$ and the PM simulation uses $N=1024^2$ particles.

Figure~\ref{fig:landau} shows the distribution function at $t=5$ and $t=25$. At $t=5$ two spiral arms have developed, which wrap up a number of times by $t=25$.

\begin{figure}
\centering
\setlength{\tabcolsep}{0pt}
\begin{tabular}{cc}
$\Delta t =\Delta x / \Delta v$ & 
$\Delta t = (1/5) \Delta x / \Delta v$ \\
\includegraphics[width=0.2\textwidth]{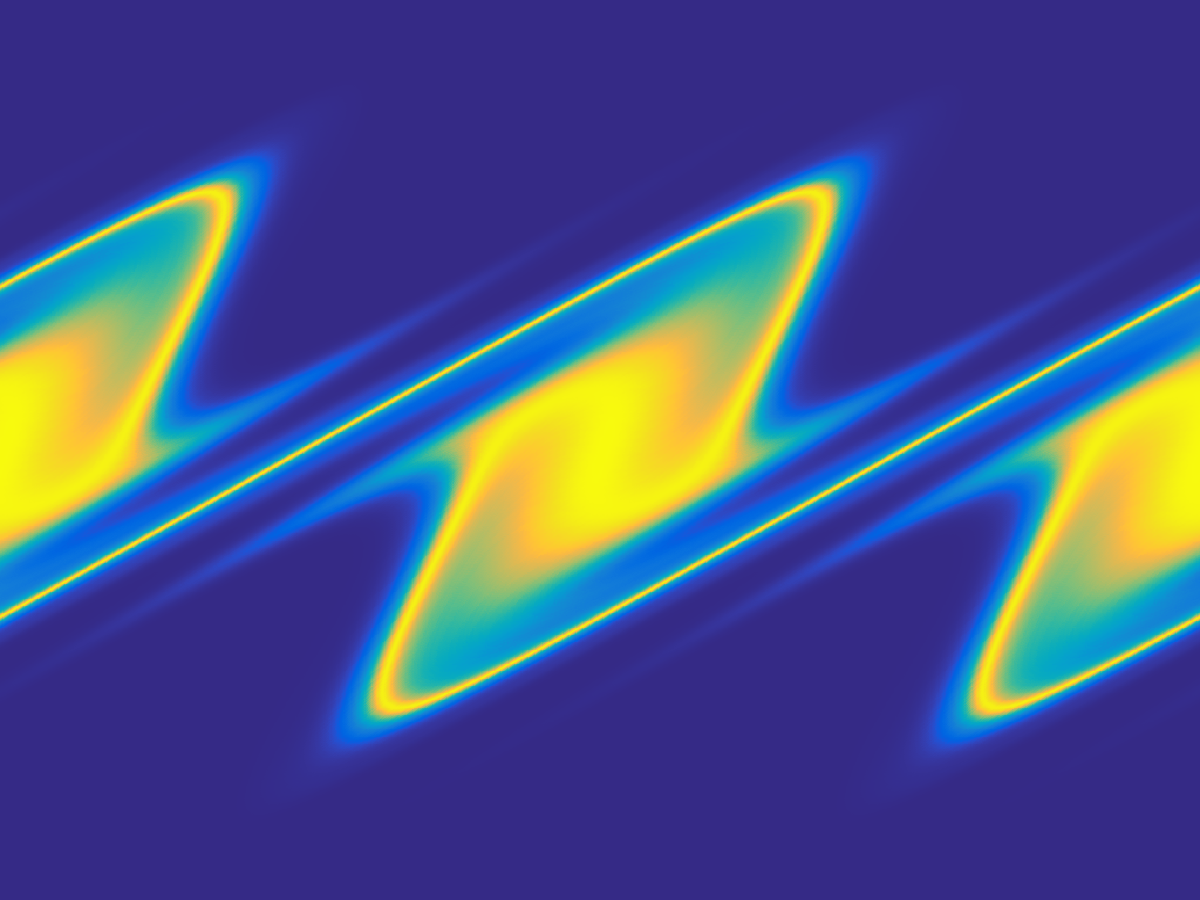}  & 
\includegraphics[width=0.2\textwidth]{figures/jeansIL3.png} \\
\end{tabular}
\caption{
The Jeans instability at $t=3$ simulated with IL with two different choices of time-steps.
Setting $\Delta x = \Delta v = \Delta t = 1$ (left panel), following the original formulation of IL, may be insufficient to resolve the dynamical timescale of the problem. This can be fixed by reducing the time-step (right panel).}
\label{fig:DT}
\end{figure}

\renewcommand{\arraystretch}{2}
\begin{table*}
\begin{tabular}{cccccccccc}
method & 
conservative? & 
reversible? & 
\shortstack{Galilean- \\ invariant?} & 
error &
resolution & 
\shortstack{computational \\  cost [$\mathcal{O}(\cdot)$]} & 
\shortstack{memory \\ scaling} & 
\shortstack{computing time \\ for test~\ref{sec:jeans} [s]}
\\
\hline 
\hline\\[-4ex]
IL & \checkmark & \checkmark & \checkmark & `lattice noise' & $N_x^3N_v^3$ & \shortstack{$N_tN_x^3N_v^3$ \\ $N_t^2N_x^3N_v^3\,^*$} & \shortstack{$N_x^3N_v^3$ \\ $N_tN_x^3\,^*$} & \shortstack{45 \\ 3} \\
PM & \checkmark & to round-off & \checkmark & Monte-Carlo & $N$ & $N_tN$ & $N$ & 10 \\
FV & \checkmark & $\times$ & $\times$ & $2^{\rm nd}$-order & $N_x^3N_v^3$ & $N_tN_x^3N_v^3$ & $N_x^3N_v^3$ & 800\\
MM & \checkmark & $\times$ & \checkmark & $2^{\rm nd}$-order & $N_x^3N_v^3$ & $N_tN_x^3N_v^3$ & $N_x^3N_v^3$ & 1000\\
\hline
{\scriptsize $^*$  memory efficient version} & & & & & & & \\
\end{tabular}
\caption{Overview and comparison of the numerical methods presented in this paper. The computing time is calculated for the time it takes to evolve test~\ref{sec:jeans} to time $t=3$, which requires $30$ time-steps in the IL and PM methods and $\sim 10000$ time-steps in the FV and MM methods.}
\label{tbl:effic}
\end{table*}

\section{Discussion}\label{sec:disc}

We compare the results of IL, PM, FV, and MM at the same number of resolution elements.
IL is seen to give accurate results.

In the Jeans instability test (Section~\ref{sec:jeans}), IL is able to resolve the thin stream feature in phase-space at $t=3$. PM resolves the feature with some Monte-Carlo noise associated with its width. FV adds significant thickness to the stream due to numerical diffusion, which is mitigated to a good extent with the MM improvement of the scheme. 
At $t=9$, the phase-space density structure becomes more intricate. Both IL and PM resolve similar small-scale features, with some associated `lattice noise' and Monte Carlo noise respectively. However, the features are washed out at this resolution with the FV and MM methods due to numerical diffusion. Again, MM shows less diffusion and sharper features than FV.

In the Gaussian Landau damping test (Section~\ref{sec:landau}), IL does a great job resolving the thin spiral arm structure at $t=5$ and $t=25$. The spiral arms are expected to wind up into thinner and thinner arms, and the solution approaches close to (but not exactly the same as) a smooth thermal equilibrium solution \citep{2005MNRAS.359..123A,2014MNRAS.441.2414C}. 
The associated diffusion with the FV smears out the thin arms, and the solution approaches the smooth, steady-state thermal equilibrium solution (i.e., making the method non-reversible). FV again shows significantly more diffusion than MM: at time $t=5$, even the large-scale structure is affected by FV's numerical diffusion as the spiral structure is not yet converged to the correct maximal velocity. 

Setting the time-step small enough to resolve the dynamical timescale is important for the IL approach, one does not simply set $\Delta t = \Delta x/\Delta v$ in general.
Figure~\ref{fig:DT} illustrates this: it re-simulates the Jeans instability with a $\Delta t = \Delta x/\Delta v$ time-step and shows the solution does not resolve the correct dynamics. This is because for our choice of resolution ($N_x$ and $N_v$) the time $\Delta x/\Delta v$ is always larger than the dynamical timescale in this problem.

All our simulations use the same memory requirements. With this condition for comparison, IL is computationally the most efficient, due to its simplicity (Table~\ref{tbl:effic}). 

We have seen that numerical diffusion in FV can alter the solution in phases-space and even relax it to a steady-state solution, losing reversibility. The results can be improved with higher order methods with less diffusive numerical fluxes, and also using a moving mesh (MM).

\subsection{Computational Efficiency}\label{sec:effic}

Table~\ref{tbl:effic} summarizes the different methods, and shows the computational and memory costs. IL is the most efficient per fixed memory cost. Due to its simplicity, our implementation of IL took about $1/3$ the compute time as PM. Both methods can be evolved on similar time-steps, since they are Lagrangian schemes. Our implementation of FV and MM took about two orders of magnitude longer to compute, primarily due to the reduction of the time-step needed for numerical stability in a not fully-Lagrangian scheme. 

Table~\ref{tbl:effic} also summarizes how much memory can be saved with a memory-efficient implementation of IL.

\subsection{Extensions}\label{sec:extension}

Here we describe possible extensions of IL for future work. IL may be applied to other types of systems to gain insight into the full phase-space distributions. One such application may be the Vlasov-Maxwell equations for collisionless plasma dynamics, where currently particle-in-cell (PIC) codes are most effective, but face the challenges of Monte-Carlo sampling noise. 

A second extension may be to simulate the weakly-collisional regime, where typically direct simulation Monte-Carlo (DSMC) techniques are employed \citep{2014MNRAS.438.2995W}. With IL, a collision operator could be developed to follow the `kick' and `drift' steps to redistribute the velocities at each physical location. The technique may be able to provide valuable insight into the ram-pressure stripping of galaxies, which is physically in the weakly-collisional regime but typically simulated in the collisional fluid regime due to computational expense.

\section{Concluding Remarks}\label{sec:conc}

The IL scheme is an easy and efficient method to directly simulate the entire phase-space of a collisionless, self-gravitating fluid (such as dark matter). It is competitive with other methods such as PM, FV, and MM, and may be the method of choice to gain insight into certain physical systems. Uniquely, IL is exactly reversible, unaffected by round-off. No doubt, it may be used to gain quick insight into full phase-space evolution as it is the fastest method per fixed memory requirement. Due to its simplicity, it is possible to reduce the memory requirements of IL from $N^6$ (a critical limitation of direct phase-space integration methods) to $N^4$, making it useful for future investigation of fine-grained phase-space evolution of collisionless dark matter and stellar systems.

\section*{Acknowledgements} This material is based upon work supported 
by the National Science Foundation Graduate Research Fellowship under 
grant no. DGE-1144152. PM is supported in part by the NASA Earth and 
Space Science Fellowship.  
The authors would like to thank Scott Tremaine for critical reading
of the manuscript and valuable suggestions.

\bibliography{mybib}{}

\bsp
\label{lastpage}
\end{document}